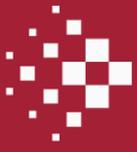
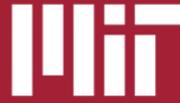

# Mapping AI Risk Mitigations

Evidence Scan and Preliminary Mitigation Taxonomy

December 2025

Alexander K. Saeri, Sophia Lloyd George, Jess Graham, Clelia Lacarriere, Peter Slattery, Michael Noetel, Neil Thompson

# Mapping AI Risk Mitigations: Evidence Scan and Preliminary AI Risk Mitigation Taxonomy

Alexander K. Saeri[1,2,*]    Sophia Lloyd George[1,3]    Jess Graham[2]
Clelia D. Lacarriere[1]    Peter Slattery[1]    Michael Noetel[2]    Neil Thompson[1]

[1]MIT FutureTech    [2]The University of Queensland
[3]Cambridge Boston Alignment Initiative

## Abstract

Organizations and governments that develop, deploy, use, and govern AI must coordinate on effective risk mitigation. However, the landscape of AI risk mitigation frameworks is fragmented, uses inconsistent terminology, and has gaps in coverage. This paper introduces a preliminary AI Risk Mitigation Taxonomy to organize AI risk mitigations and provide a common frame of reference. The Taxonomy was developed through a rapid evidence scan of 13 AI risk mitigation frameworks published between 2023–2025, which were extracted into a living database of 831 distinct AI risk mitigations. The mitigations were iteratively clustered & coded to create the Taxonomy. The preliminary AI Risk Mitigation Taxonomy organizes mitigations into four categories: (1) Governance & Oversight: Formal organizational structures and policy frameworks that establish human oversight mechanisms and decision protocols; (2) Technical & Security: Technical, physical, and engineering safeguards that secure AI systems and constrain model behaviors; (3) Operational Process: processes and management frameworks governing AI system deployment, usage, monitoring, incident handling, and validation; and (4) Transparency & Accountability: formal disclosure practices and verification mechanisms that communicate AI system information and enable external scrutiny. These categories are further subdivided into 23 mitigation subcategories. The rapid evidence scan and taxonomy construction also revealed several cases where terms like 'risk management' and 'red teaming' are used widely but refer to different responsible actors, actions, and mechanisms of action to reduce risk. This Taxonomy and associated mitigation database, while preliminary, offers a starting point for collation and synthesis of AI risk mitigations. It also offers an accessible, structured way for different actors in the AI ecosystem to discuss and coordinate action to reduce risks from AI.

[*] Corresponding author: aksaeri@mit.edu

# 1   Introduction

To address risks from increasingly capable Artificial Intelligence (AI), effective mitigations must be developed and implemented. For this task, many actors - from researchers to industry leaders - must be able to coordinate action and communicate clearly about AI risk mitigations.

However, as awareness and concerns of AI risks has increased (Center for AI Safety, 2023; Bengio et al., 2025), the field has become more fragmented and less coordinated (Slattery et al., 2024). Organizations that develop, deploy, use, and govern AI have generated a variety of proposed mitigations, safeguards, and governance mechanisms to address risks (e.g., NIST, 2024; Eisenberg, 2025). Frameworks, standards, and other documents approach mitigations from different disciplinary or practice backgrounds, use diverging terminology, different theories, and inconsistent classifications. Some focus on adapting established mitigations from cybersecurity or safety-critical industries (e.g., incident response, system shutdown; Koessler & Schuett, 2023), while others introduce novel approaches specific to AI (e.g., alignment techniques, model interpretability; Ji et al., 2023). The result is a proliferation of overlapping, incomplete, and sometimes incompatible mitigation frameworks.

This fragmented landscape has theoretical and practical consequences. A lack of shared definitions and structures makes incremental scientific progress challenging. The reinvention and duplication also lead to fragmentation and confusion. For example, 'red teaming' can include many different methods, to evaluate many different threat models, and little consensus on who should perform it (Feffer, 2024). Without an accessible or pragmatic shared understanding of risk mitigations, the actors struggle to develop, implement and coordinate mitigations. As noted by the U.S.–EU Trade and Technology Council in its Joint Roadmap for Trustworthy AI and Risk Management, "shared terminologies and taxonomies are essential for operationalizing trustworthy AI and risk management in an interoperable fashion" (European Commission and the United States Trade and Technology Council, 2022).

These challenges are compounded by the rapid and accelerating pace of AI development and adoption. The share of organizations using AI in at least one business function quadrupled from 20% in 2017 to 80% in 2024 (Singla et al., 2024). The adoption of highly capable general-purpose AI agents tripled between Q1 (11%) and Q2 (33%) 2025 alone (KPMG, 2025). This expansion significantly increases the number of stakeholders who must implement mitigations. It also increases the diversity of contexts in which effective risk management must occur.

To address this gap, we conducted an evidence scan of public AI risk mitigation frameworks, with the aim of identifying, extracting, and systematizing mitigations across policy, technical, and risk management reports. We used methods adapted from evidence synthesis approaches (Khangura, 2012) and framework synthesis approaches (Carroll et al., 2011; 2013) to identify and extract mitigations into a publicly accessible AI Risk Mitigation Database. These mitigations were then iteratively clustered and categorized to construct a preliminary AI Risk Mitigation Taxonomy. This work was undertaken as part of a broader AI Risk Initiative (airisk.mit.edu), which aims to help decision-makers identify, prioritize, and manage risks from AI.

The major contribution of this work is in creating a common frame of reference for AI risk mitigations. Both the Database and the Taxonomy are released publicly on the AI Risk Initiative website (airisk.mit.edu) for iteration, feedback, and use because (1) we observe growing demand for a comprehensive, accessible synthesis of AI mitigations from diverse sources, and (2) shared understanding is a precondition for effective risk management.

The preliminary AI Risk Mitigation Database and Taxonomy together provide an empirical and conceptual foundation for a more coordinated, comprehensive approach to mitigating AI risks. They are intended to support a wide range of actors and stakeholders in identifying, developing, implementing, and coordinating action to address risks from AI.

## 2 Methods

### 2.1 Definitions

- We define **artificial intelligence (AI)** as "systems or machines capable of performing tasks that typically require human intelligence" (Bengio et al., 2025)

- We define **AI risk** as "the possibility of an unfortunate occurrence that may emerge from the development, deployment or use of AI" after the Society for Risk Analysis (Aven et al., 2018).

- We define **mitigation** as "an action that reduces the likelihood or impact of a risk" after the Society for Risk Analysis (Aven et al., 2018) and other authoritative sources (Actuarial Standards Board, 2014; NIST, 2012).

We therefore define an **AI risk mitigation** as "an action that reduces the likelihood or impact of an unfortunate occurrence that may emerge from the development, deployment, or use of systems or machines capable of performing tasks that typically require human intelligence".

### 2.2 Overview of approach

Our overall approach was a rapid evidence scan, a modified type of evidence synthesis (Khangura, 2012). We identified foundational, widely discussed documents that proposed frameworks of AI risk mitigations. We extracted mitigations from those documents into a database. We iteratively analyzed, clustered, and labelled mitigations to develop a taxonomy of AI risk mitigations. We experimented with large language models (LLMs) as assistive tools, but mixed results[†] meant that extracting and classifying mitigations was ultimately conducted by human authors.

### 2.3 Document identification and selection

We used a purposive sampling strategy to identify public documents that presented structured frameworks of AI risk mitigations, published in the period 2023-2025. We took an initial set of eligible documents and expanded the list by (1) backwards reference mining from the initial set and (2) searching for works by high-profile authors and organizations who work on AI risk mitigations. Potentially relevant documents were informally screened by the authorship team for eligibility.

---

[†] Consistent with emerging practice in evidence synthesis (Nguyen-Trung et al., 2024; Cao et al., 2025), we evaluated large language models (LLMs) as assistive tools during extraction and classification. LLMs were used to identify potentially relevant mitigations from documents, and recommend and justify classifications against the iterated taxonomy. We implemented a quality assurance process when using LLMs as assistive tools: (1) a document-level review where LLM-extracted mitigations for each document were compared with an author's manual extraction of the same mitigations; (2) a classification audit where LLM-recommended classifications & justifications for each mitigation were reviewed by an author and accepted or changed; and (3) cross-checking by multiple authors to assess inter-annotator consistency. The quality assurance process found that LLMs could not reliably be used to automate extraction; they would sometimes confabulate, combine, or omit mitigations. LLMs were more useful when recommending classification against the taxonomy, because all classifications were reviewed by human authors. These results suggest that rigorous scaffolding and human validation is needed to benefit LLMs as assistive tools in evidence synthesis research.

We included a total of 13 foundational, widely discussed documents from governments, standards bodies, and governance or research organizations (see Table 2). Each document proposed a framework or other structured list of AI risk mitigations.

## 2.4 Mitigation extraction into preliminary AI Risk Mitigation Database

We extracted 831 distinct mitigations from the included documents into a structured database using a standardized coding frame. The database included: (1) a unique mitigation identifier; (2) a mitigation name (verbatim when available, otherwise constructed by the authors); and (3) a description of the mitigation (verbatim when available, otherwise excerpted by the authors or omitted). Items were extracted only if they satisfied the definition of an AI risk mitigation. The database is publicly available at airisk.mit.edu.

## 2.5 Construction of preliminary AI Risk Mitigation Taxonomy

We developed our AI risk mitigation taxonomy using an iterative approach that combined thematic synthesis (Thomas & Harden, 2008) and framework synthesis (Ritchie & Spencer, 2002). Thematic synthesis is a "bottom up" method where concepts (i.e., mitigation names & definitions) are iteratively analyzed to identify patterns and structure. Framework synthesis is a "top down" method where concepts are coded against a pre-existing structure. The approach we used is similar to a "best fit" framework synthesis approach (Carroll et al., 2011; 2013) except that we did not select any single framework identified in the evidence scan as a "best" initial framework to iterate upon. Figure 1 summarizes the process.

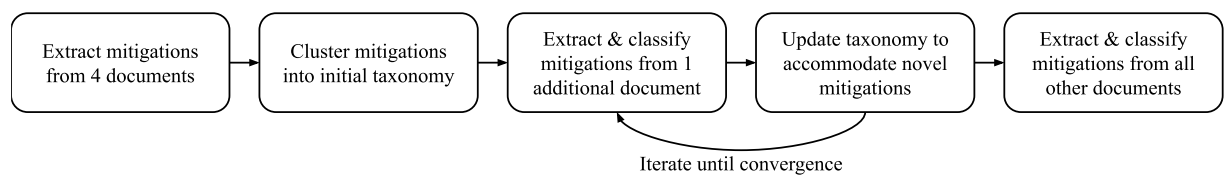

Figure 1. Overview of process to construct preliminary AI Risk Mitigation Taxonomy

We first manually extracted mitigations from 4 documents, and clustered mitigations from those documents using thematic analysis. We experimented with several structures when we conducted the initial clustering, including:

- Risk management stage: identification, assessment, mitigation, etc. (e.g., NIST, 2012; 2023)

- AI system lifecycle: Design, development (training, testing), deployment, and use (e.g,. Kangeter, 2024)

- Actor-based approach: who designs, implements, enforces, or is affected by mitigations (e.g., NIST, 2023; Srikumar et al., 2024)

- Risk-based: clustering by the specific risk being addressed by a mitigation (e.g., as described in the MIT AI Risk Taxonomy; Slattery et al., 2024)

- Technical vs. socio-technical: clustering by whether the intervention or target of a mitigation was the technical AI model or human behavior (e.g., Weidinger et al., 2023)

We found that using elements of the *AI system lifecycle* and *socio-technical structure* accommodated the largest proportion of mitigations. This combined structure was used as an initial taxonomy. After we classified all the extracted mitigations according to the initial taxonomy, we then identified mitigations that could not be classified and gathered internal feedback from our authorship team. We modified the taxonomy to accommodate any unclassified mitigations, then tested the updated taxonomy on mitigations from a new document. We repeated this process three times in total.

Over three iterations, we converged on a taxonomy that clustered mitigations into four categories and 23 subcategories. All mitigations from the documents were then classified using this taxonomy. We extracted 831 mitigations and classified 815 mitigations (98%) using the taxonomy. Sixteen mitigations could not be classified.

## 3 Results

### 3.1 Preliminary AI Risk Mitigation Taxonomy

The preliminary AI Risk Mitigation Taxonomy is presented in Table 1. This comprises four categories: Governance & Oversight, Technical & Security, Operational Process, and Transparency & Accountability and 23 subcategories. The Taxonomy is reproduced with example mitigations for each subcategory in Appendix A.

Table 1. Preliminary AI Risk Mitigation Taxonomy

**1. Governance & Oversight Mitigations**
*Formal organizational structures and policy frameworks that establish human oversight mechanisms and decision protocols to ensure human accountability, ethical conduct, and risk management throughout AI development and deployment*

| | |
|---|---|
| 1.1 Board Structure & Oversight | Governance structures and leadership roles that establish executive accountability for AI safety and risk management. |
| 1.2 Risk Management | Systematic methods that identify, evaluate, and manage AI risks for comprehensive risk governance across organizations. |
| 1.3 Conflict of Interest Protections | Governance mechanisms that manage financial interests and organizational structures to ensure leadership can prioritize safety over profit motives in critical situations. |
| 1.4 Whistleblower Reporting & Protection | Policies and systems that enable confidential reporting of safety concerns or ethical violations to prevent retaliation and encourage disclosure of risks. |
| 1.5 Safety Decision Frameworks | Protocols and commitments that constrain decision-making about model development, deployment, and capability scaling, and govern safety-capability resource allocation to prevent unsafe AI advancement. |
| 1.6 Environmental Impact Management | Processes for measuring, reporting, and reducing the environmental footprint of AI systems to ensure sustainability and responsible resource use. |
| 1.7 Societal Impact Assessment | Processes that assess AI systems' effects on society, including impacts on employment, power dynamics, political processes, and cultural values. |

## 2. Technical & Security Mitigations

*Technical, physical, and engineering safeguards that secure AI systems and constrain model behaviors to ensure security, safety, alignment with human values, and content integrity.*

| | |
|---|---|
| 2.1 Model & Infrastructure Security | Technical and physical safeguards that secure AI models, weights, and infrastructure to prevent unauthorized access, theft, tampering, and espionage. |
| 2.2 Model Alignment | Technical methods to ensure AI systems understand and adhere to human values and intentions. |
| 2.3 Model Safety Engineering | Technical methods and safeguards that constrain model behaviors and protect against exploitation and vulnerabilities. |
| 2.4 Content Safety Controls | Technical systems and processes that detect, filter, and label AI-generated content to identify misuse and enable content provenance tracking. |

## 3. Operational Process Mitigations

*Processes and management frameworks governing AI system deployment, usage, monitoring, incident handling, and validation, which promote safety, security, and accountability throughout the system lifecycle*

| | |
|---|---|
| 3.1 Testing & Auditing | Systematic internal and external evaluations that assess AI systems, infrastructure, and compliance processes to identify risks, verify safety, and ensure performance meets standards. |
| 3.2 Data Governance | Policies and procedures that govern responsible data acquisition, curation, and usage to ensure compliance, quality, user privacy, and removal of harmful content. |
| 3.3 Access Management | Operational policies and verification systems that govern who can use AI systems and for what purposes to prevent safety circumvention, deliberate misuse, and deployment in high-risk contexts. |
| 3.4 Staged Deployment | Implementation protocols that deploy AI systems in stages, requiring safety validation before expanding user access or capabilities. |
| 3.5 Post-Deployment Monitoring | Ongoing monitoring processes that track AI behavior, user interactions, and societal impacts post-deployment to detect misuse, emergent dangerous capabilities, and harmful effects. |
| 3.6 Incident Response & Recovery | Protocols and technical systems that respond to security incidents, safety failures, or capability misuse to contain harm and restore safe operations. |

## 4. Transparency & Accountability Mitigations

*Formal disclosure practices and verification mechanisms that communicate AI system information and enable external scrutiny to build trust, facilitate oversight, and ensure accountability to users, regulators, and the public.*

| | |
|---|---|
| 4.1 System Documentation | Comprehensive documentation protocols that record technical specifications, intended uses, capabilities, and limitations of AI systems to enable informed evaluation and governance. |
| 4.2 Risk Disclosure | Formal reporting protocols and notification systems that communicate risk information, mitigation plans, safety evaluations, and significant AI activities to enable external oversight and inform stakeholders. |
| 4.3 Incident Reporting | Formal processes and protocols that document and share AI safety incidents, security breaches, near-misses, and relevant threat intelligence with appropriate stakeholders to enable coordinated responses and systemic improvements. |
| 4.4 Governance Disclosure | Formal disclosure mechanisms that communicate governance structures, decision frameworks, and safety commitments to enhance transparency and enable external oversight of high-stakes AI decisions. |
| 4.5 Third-Party System Access | Mechanisms granting controlled system access to vetted external parties to enable independent assessment, validation, and safety research of AI models and capabilities. |
| 4.6 User Rights & Recourse | Frameworks and procedures that enable users to identify and understand AI system interactions, report issues, request explanations, and seek recourse or remediation when affected by AI systems. |

### 3.2 Characteristics of included documents

We included 13 foundational documents from a variety of sources which proposed structured frameworks or taxonomies of AI risk mitigations, published between 2023 and 2025. Documents were primarily published from institutions or authors in the US, EU or UK, and included government research reports, academic research reports, standards guidance, and proposed legislation. Some of the documents explicitly mapped their proposed mitigations to risk categories (e.g., Eisenberg et al., 2025 mapped CONTROL-037, "implement AI alignment validation system" to the risk "AI pursuing its own goals in conflict with human goals or values"). Other documents implicitly mapped mitigations to risks by including mention of risk(s) in mitigation definitions (e.g., DSIT UK, 2023 included the mitigation "Model Evaluation of Dangerous Capabilities"). Finally, some documents did not include any explicit or implicit mapping of mitigations to risks. The title, first author, number of mitigations extracted, Risk mapping type, and year of publication are presented in Table 2.

Table 2. Characteristics of included documents.

| Title | First Author | Mitigations | Mitigation-Risk Mapping | Year |
| --- | --- | --- | --- | --- |
| International AI Safety Report 2025 | Bengio, Y. | 26 | None | 2025 |
| A Frontier AI Risk Management Framework | Campos, S. | 20 | None | 2025 |
| The Unified Control Framework | Eisenberg, I. | 42 | Explicit | 2025 |
| Risk Sources and Risk Management Measures in support of standards for general-purpose AI systems | Gipiškis, D. | 83 | Explicit | 2024 |
| Effective Mitigations for Systemic Risks from General-Purpose AI | Uuk, R. | 28 | Explicit | 2024 |
| FLI AI Safety Index 2024 | Future of Life Institute | 87 | Explicit | 2024 |
| Towards Best Practices in AGI Safety and Governance | Schuett, J. | 51 | Implicit | 2023 |
| Pitfalls of Evidence-Based AI Policy | Casper, S. | 15 | None | 2025 |
| EU AI Act: General Purpose AI Code of Practice (Draft 3) | EU AI Office | 16 | Explicit | 2025 |
| Emerging Processes for Frontier AI Safety | Department for Science, Innovation and Technology (UK) | 170 | Implicit | 2023 |
| NIST AI Risk Management Framework: Generative AI Profile | National Institute of Standards and Technology (USA) | 203 | Explicit | 2024 |
| AI Risk Management Standards Profile for GPAIS | Barrett, A. M. | 80 | Explicit | 2024 |
| California Senate Bill 1047 | California State Legislature | 10 | None | 2024 |

The distribution of mitigations across each of our taxonomy categories and subcategories is shown in Table 3.

Table 3. Classification of extracted AI risk mitigations

| Category / Subcategory | | Classified Mitigations | | Documents |
|---|---|---|---|---|
| | | n | % | |
| *1* | *Governance & Oversight Controls* | *248* | *30%* | *13* |
| 1.1 | Board Structure & Oversight | 35 | 4% | 8 |
| 1.2 | Risk Management | 125 | 15% | 13 |
| 1.3 | Conflict of Interest Protections | 8 | 1% | 3 |
| 1.4 | Whistleblower Reporting & Protection | 10 | 1% | 7 |
| 1.5 | Safety Decision Frameworks | 31 | 4% | 11 |
| 1.6 | Environmental Impact Management | 11 | 1% | 5 |
| 1.7 | Societal Impact Assessment | 28 | 3% | 7 |
| *2* | *Technical & Security Controls* | *101* | *12%* | *12* |
| 2.1 | Model & Infrastructure Security | 32 | 4% | 11 |
| 2.2 | Model Alignment | 9 | 1% | 5 |
| 2.3 | Model Safety Engineering | 38 | 5% | 7 |
| 2.4 | Content Safety Controls | 22 | 3% | 6 |
| *3* | *Operational Process Controls* | *295* | *36%* | *13* |
| 3.1 | Testing & Auditing | 127 | 15% | 12 |
| 3.2 | Data Governance | 57 | 7% | 7 |
| 3.3 | Access Management | 23 | 3% | 7 |
| 3.4 | Staged Deployment | 8 | 1% | 5 |
| 3.5 | Post-deployment Monitoring | 50 | 6% | 9 |
| 3.6 | Incident Response & Recovery | 30 | 4% | 9 |
| *4* | *Transparency & Accountability Controls* | *171* | *21%* | *13* |
| 4.1 | System Documentation | 37 | 4% | 10 |
| 4.2 | Risk Disclosure | 44 | 5% | 11 |
| 4.3 | Incident Reporting | 30 | 4% | 11 |
| 4.4 | Governance Disclosure | 24 | 3% | 9 |
| 4.5 | Third-Party System Access | 19 | 2% | 7 |
| 4.6 | User Rights & Recourse | 19 | 2% | 5 |
| *X.X* | *Mitigation not otherwise categorized* | *16* | *2%* | *8* |

Mitigations were distributed unevenly across the four high-level categories. The largest share fell under Operational Process Controls (n = 295, 36%), which primarily included testing and auditing (n = 127), data governance (n = 57), and post-deployment monitoring (n = 50). The next largest group comprised Governance & Oversight Controls (n = 248, 30%), including risk management (n = 125), board structure and oversight (n = 35), and safety decision frameworks (n = 31). Transparency & Accountability Controls accounted for n = 171 (21%), with common subcategories such as risk disclosure (n = 44), system documentation (n = 37), and incident reporting (n = 30). Finally, Technical & Security Controls comprised n = 101 (12%), primarily model and infrastructure security (n = 32) and model safety engineering (n = 38).

The 16 mitigations that could not be classified (2%) were distributed across 8 of 13 documents. These uncategorized mitigations were primarily quality assurance actions (e.g., "implement scalable AI infrastructure", "AI system lifecycle management") or organizational strategy and communications actions (e.g., "avoiding hype", "testimonies to policymakers"). Some mitigations described broad organizational conditions such as "safety culture" or "company risk culture" that do not easily translate into discrete, implementable actions. The full list of uncategorized mitigations is presented in Appendix B.

## 4 Discussion

Across 13 foundational documents published between 2023 and 2025, we extracted 831 distinct AI risk mitigations and organized them into a preliminary AI Risk Mitigation Taxonomy comprising four categories and 23 subcategories. This taxonomy provides a structured foundation for identifying, comparing, and evaluating AI risk mitigations: actions that reduce the likelihood or impact of risks from the development, deployment, or use of AI.

### 4.1 Insights from the included literature

#### 4.1.1 Risk Management is widely discussed but inconsistently defined

We find that *1.2 Risk Management* is one of the most widely referenced subcategories of AI mitigations in the included documents. All 13 documents referred to risk management, and 125 mitigations (15% of total) were classified as risk management according to the preliminary taxonomy: "Systematic methods that identify, evaluate, and manage AI risks for comprehensive risk governance across organizations".

However, the term risk management is inconsistently defined and operationalized across the included documents. Some documents provided structured definitions (e.g., NIST, 2024; Campos et al., 2025; Gipiškis et al., 2024; Bengio et al., 2025), but these definitions varied in scope and specificity. Where provided, definitions typically included several related stages or functions, including identifying risks, evaluating risks, and implementing measures to reduce risks. Some definitions also emphasized the role of governance - rules, procedures, or culture - to provide structure the other functions (e.g., NIST, 2024); others emphasized monitoring AI system behavior once it is deployed and used (e.g., Eisenberg et al., 2025). Indeed, Schuett et al. (2023) observed that, from a survey of AI experts, "enterprise risk management" was the practice with the highest "I don't know" response (26%); as Schuett et al. remark, this "indicates that many respondents simply did not know what enterprise risk management is and how it works" (p. 12).

We propose that this reflects that 'AI risk management' is an emerging concept; the boundaries of risk management are not yet settled (i.e., what should be included or excluded as a risk management action). The contested definitions of risk management also extend to our preliminary taxonomy. For example, other mitigation subcategories have conceptual overlap with the specific functions of risk management (e.g., identifying, analyzing, enacting measures, governing and/or monitoring), like safety decision frameworks (1.5), risk disclosure (4.2), and post-deployment monitoring (3.5).

### 4.1.2 Common and uncommon categories of AI risk mitigations

We found that *3.1 Testing & Auditing* was the most frequently referenced subcategory of AI mitigations. 12 of 13 documents referred to this subcategory and 127 mitigations (15%) were classified as Testing & Auditing: "Systematic internal and external evaluations that assess AI systems, infrastructure, and compliance processes to identify risks, verify safety, and ensure performance meets standards". Other existing mitigations subcategories have conceptual overlap with testing & auditing actions. Risk management (1.2) involves testing via risk assessments that evaluate the performance of an AI model or system. Model safety engineering (2.3) can involve safety analysis protocols and hierarchical auditing. Third-party system access (4.5) lets researchers or governments access models for pre- or post-deployment testing.

Several mitigation subcategories were much less frequently mentioned in the included documents. These categories include *Conflict of Interest Protections, Whistleblower Reporting & Protection, Environmental Impact Management, Model Alignment, and Staged Deployment*. Each represented <1% of distinct mitigations identified in this evidence scan. Several of these categories of mitigations were also relatively underrepresented across included documents. For example, *Conflict of Interest Protections* appeared in only 3 of 13 documents. *Model Alignment, Environmental Impact Management, Staged Deployment*, and *User Rights & Recourse* appeared in only 5 of 13 documents. Of course, the frequency with which a mitigation is classified does not necessarily reflect its relative importance, or its rate of adoption, but further work should determine whether these mitigations are merely uncommonly mentioned or in fact neglected.

### 4.2 Practical implications

While preliminary, our AI Risk Mitigation Taxonomy and associated Database can support risk-management practices across multiple stakeholder groups. For *Technology Managers and AI Developers*, the taxonomy can help navigate and assess organizational capacity in development and implementation of AI risk mitigations. For *Policymakers and Regulators* it can provide specific, actionable categories of mitigations to guide policy and oversight. For instance, regulators can use the taxonomy to translate general obligations (e.g., "implement appropriate risk controls") into more measurable and precise criteria. Similarly, *Auditors and Compliance Officers* can use the taxonomy to decompose comprehensive AI assurance processes into more specific areas of interest. This can support more systematic and coherent evaluation across technical, organizational, and governance dimensions.

### 4.3 Future directions for research

A critical next step is to map AI risk mitigations to the risks they are intended to address. Many of the mitigations identified in this study do not specify which risks they are intended to address, which makes selection, implementation, and evaluation more difficult. Comprehensive mapping would facilitate integrated risk governance frameworks. For example, Gipiškis et al. (2024) propose linking "risk sources" to "risk measures," an approach that could be expanded through integration with emerging standards (e.g., NIST, 2024). Such mappings would enable more precise identification of strengths and gaps in risk management. It would help practitioners understand which mitigations are relevant for specific categories of risk.

We also suggest researchers explore organizational conditions that reduce AI risks but are not easily captured as discrete actions. For instance, Campos et al. (2025) emphasize "safety culture" as a core component of frontier AI risk governance. Other organizational factors, such as competitive dynamics, psychological safety, and team structures, may influence the effectiveness of formal mitigation measures. Understanding these sociotechnical determinants of safety could help explain why some organizations implement mitigations effectively while others do not. The effectiveness of a mitigation is dependent on the quality of its implementation.

Our evidence scan focused on foundational, widely discussed documents (e.g., Bengio et al., 2025). These documents described mitigations relevant to organizations developing or deploying the most capable general-purpose AI systems ('frontier AI'). In this set of documents, there was comparatively less attention toward AI systems, use cases, and mitigations relevant to other actors—such as regulators, purchasers, users, or affected communities. Future research could examine how mitigations differ across actor types, and how their actions interact. A more holistic view of the AI risk ecosystem could improve coordination and accountability by exploring how interventions across actors reinforce or counteract each other.

### 4.4 Limitations

Our rapid evidence scan did not include a systematic search or extensive expert consultation to identify other relevant documents. It also excluded documents not in English and did not consider mitigations for AI risks in specific sectors, or for specific use cases. The iterative, thematic approach to taxonomy construction could reproduce conceptual confusions in the literature, such as the definition of risk management, and the boundary of testing & auditing. Finally, while our taxonomy captures the presence of mitigations, it does not evaluate their effectiveness, cost, or interaction effects.

These limitations could be meaningfully addressed by a systematic review of AI risk mitigations. In such a review, a wider panel of experts could help identify and classify a comprehensive set of documents to increase rigor while maintaining practicality for research end-users. These refinements could improve interoperability with other frameworks and efforts to improve coordination and action on AI risk (e.g., Slattery et al., 2024; Bagehorn et al., 2025).

## 5 Conclusion

In this research, we conducted a rapid systematic evidence scan that identified 13 foundational documents that proposed AI risk mitigation frameworks. We extracted and classified 831 distinct mitigations. These mitigations were organized into a preliminary AI Risk taxonomy with four categories: Governance & Oversight, Technical & Security, Operational Process, and Transparency & Accountability. The preliminary AI Risk Mitigation Taxonomy provides an accessible entry point for understanding the evolving landscape of actions that reduce the likelihood or impact of risks from the development, deployment, or use of AI.

# References

*Note. Bolding indicates a document included in the evidence scan.*

**Appendix A: Preliminary AI Risk Mitigation Taxonomy with Examples**

View the database and interactive taxonomy at airisk.mit.edu.

| Mitigation Category | Mitigation Subcategory | Subcategory description | Examples |
|---|---|---|---|
| **1. Governance & Oversight Mitigations**  *Formal organizational structures and policy frameworks that establish human oversight mechanisms and decision protocols to ensure human accountability, ethical conduct, and risk management throughout AI development and deployment.* | 1.1 Board Structure & Oversight | Governance structures and leadership roles that establish executive accountability for AI safety and risk management. | *Dedicated risk committees, safety teams, ethics boards, crisis simulation training, multi-party authorization protocols, deployment veto powers* |
| | 1.2 Risk Management | Systematic methods that identify, evaluate, and manage AI risks for comprehensive risk governance across organizations. | *Enterprise risk management frameworks, risk registers with capability thresholds, compliance programs, pre-deployment risk assessments, independent risk assessments* |
| | 1.3 Conflict of Interest Protections | Governance mechanisms that manage financial interests and organizational structures to ensure leadership can prioritize safety over profit motives in critical situations. | *Background checks for key personnel, windfall profit redistribution plans, stake limitation policies, protections against shareholder pressure* |
| | 1.4 Whistleblower Reporting & Protection | Policies and systems that enable confidential reporting of safety concerns or ethical violations to prevent retaliation and encourage disclosure of risks. | *Anonymous reporting channels, non-retaliation guarantees, limitations on non-disparagement agreements, external whistleblower handling services* |
| | 1.5 Safety Decision Frameworks | Protocols and commitments that constrain decision-making about model development, deployment, and capability scaling, and govern safety-capability resource allocation to prevent unsafe AI advancement. | *If-then safety protocols, capability ceilings, deployment pause triggers, safety-capability resource ratios* |
| | 1.6 Environmental Impact Management | Processes for measuring, reporting, and reducing the environmental footprint of AI systems to ensure sustainability and responsible resource use. | *Carbon footprint assessment, emission offset programs, energy efficiency optimization, resource consumption tracking* |
| | 1.7 Societal Impact Assessment | Processes that assess AI systems' effects on society, including impacts on employment, power dynamics, political processes, and cultural values. | *Fundamental rights impact assessments, expert consultations on risk domains, stakeholder engagement processes, governance gap analyses* |

| Mitigation Category | Mitigation Subcategory | Subcategory description | Examples |
|---|---|---|---|
| **2. Technical & Security Mitigations** *Technical, physical, and engineering safeguards that secure AI systems and constrain model behaviors to ensure security, safety, alignment with human values, and content integrity.* | 2.1 Model & Infrastructure Security | Technical and physical safeguards that secure AI models, weights, and infrastructure to prevent unauthorized access, theft, tampering, and espionage. | *Model weight tracking systems, multifactor authentication protocols, physical access controls, background security checks, compliance with information security standards* |
| | 2.2 Model Alignment | Technical methods to ensure AI systems understand and adhere to human values and intentions. | *Reinforcement learning from human feedback (RLHF), direct preference optimization (DPO), constitutional AI training, value alignment verification systems* |
| | 2.3 Model Safety Engineering | Technical methods and safeguards that constrain model behaviors and protect against exploitation and vulnerabilities. | *Safety analysis protocols, capability restriction mechanisms, hazardous knowledge unlearning techniques, input/output filtering systems, defense-in-depth implementations, adversarial robustness training, hierarchical auditing, action replacement* |
| | 2.4 Content Safety Controls | Technical systems and processes that detect, filter, and label AI-generated content to identify misuse and enable content provenance tracking. | *Synthetic media watermarking, content filtering mechanisms, prohibited content detection, metadata tagging protocols, deepfake creation restrictions* |

| Mitigation Category | Mitigation Subcategory | Subcategory description | Examples |
|---|---|---|---|
| **3. Operational Process Mitigations**<br><br>*Processes and management frameworks governing AI system deployment, usage, monitoring, incident handling, and validation, which promote safety, security, and accountability throughout the system lifecycle.* | 3.1 Testing & Auditing | Systematic internal and external evaluations that assess AI systems, infrastructure, and compliance processes to identify risks, verify safety, and ensure performance meets standards. | *Third-party audits, red teaming, penetration testing, dangerous capability evaluations, bug bounty programs* |
| | 3.2 Data Governance | Policies and procedures that govern responsible data acquisition, curation, and usage to ensure compliance, quality, user privacy, and removal of harmful content. | *Harmful content filtering protocols, compliance checks for data collection standards, user data privacy controls, data curation processes* |
| | 3.3 Access Management | Operational policies and verification systems that govern who can use AI systems and for what purposes to prevent safety circumvention, deliberate misuse, and deployment in high-risk contexts. | *KYC verification requirements, API-only access controls, fine-tuning restrictions, acceptable use policies, high-stakes application prohibitions* |
| | 3.4 Staged Deployment | Implementation protocols that deploy AI systems in stages, requiring safety validation before expanding user access or capabilities. | *Limited API access programs, gradual user base expansion, capability threshold assessments, pre-deployment validation checkpoints, treating model updates as new deployments* |
| | 3.5 Post-Deployment Monitoring | Ongoing monitoring processes that track AI behavior, user interactions, and societal impacts post-deployment to detect misuse, emergent dangerous capabilities, and harmful effects. | *User interaction tracking systems, capability evolution assessments, periodic impact reports, automated misuse detection, usage pattern analysis tools* |
| | 3.6 Incident Response & Recovery | Protocols and technical systems that respond to security incidents, safety failures, or capability misuse to contain harm and restore safe operations. | *Incident response plans, emergency shutdown/rollback procedures, model containment mechanisms, safety drills, critical infrastructure protection measures* |

| Mitigation Category | Mitigation Subcategory | Subcategory description | Examples |
|---|---|---|---|
| **4. Transparency & Accountability Mitigations**<br><br>*Formal disclosure practices and verification mechanisms that communicate AI system information and enable external scrutiny to build trust, facilitate oversight, and ensure accountability to users, regulators, and the public.* | 4.1 System Documentation | Comprehensive documentation protocols that record technical specifications, intended uses, capabilities, and limitations of AI systems to enable informed evaluation and governance. | *Model cards, system architecture documentation, compute resource disclosures, safety test result reports, system prompts, model specifications* |
| | 4.2 Risk Disclosure | Formal reporting protocols and notification systems that communicate risk information, mitigation plans, safety evaluations, and significant AI activities to enable external oversight and inform stakeholders. | *Publishing risk assessment summaries, pre-deployment notifications to government, reporting large training runs, disclosing mitigation strategies, notifying affected parties* |
| | 4.3 Incident Reporting | Formal processes and protocols that document and share AI safety incidents, security breaches, near-misses, and relevant threat intelligence with appropriate stakeholders to enable coordinated responses and systemic improvements. | *Cyber threat intelligence sharing networks, mandatory breach notification procedures, incident database contributions, cross-industry safety reporting mechanisms, standardized near-miss documentation protocols* |
| | 4.4 Governance Disclosure | Formal disclosure mechanisms that communicate governance structures, decision frameworks, and safety commitments to enhance transparency and enable external oversight of high-stakes AI decisions. | *Published safety and/or alignment strategies, governance documentation, safety cases, model registration protocols, public commitment disclosures* |
| | 4.5 Third-Party System Access | Mechanisms granting controlled system access to vetted external parties to enable independent assessment, validation, and safety research of AI models and capabilities. | *Researcher access programs, third-party capability assessments, government access provisions, legal safe harbors for public interest evaluations* |
| | 4.6 User Rights & Recourse | Frameworks and procedures that enable users to identify and understand AI system interactions, report issues, request explanations, and seek recourse or remediation when affected by AI systems. | *User reporting channels, appeal processes, explanation request systems, remediation protocols, content verification* |

**Appendix B: Uncategorized Mitigations**

Here we present the 16 mitigations from the evidence scan that could not be classified using the preliminary AI Risk Mitigation Taxonomy.

| Mitigation Name | Mitigation Description | Author (Year) |
| --- | --- | --- |
| AI governance institutes | National governments can create AI governance institutes to research risks, evaluate systems, and curate best practices | Casper & Krueger (2025) |
| AI system integration framework | Framework for AI system integration including architecture review, compatibility testing, and validation | Eisenberg et al (2025) |
| AI system lifecycle management | Systematic processes for maintenance, updates, and retraining including version control and deployment pipelines | Eisenberg et al (2025) |
| Scalable AI infrastructure | Architecture and infrastructure practices to ensure AI systems can scale effectively | Eisenberg et al (2025) |
| AI literacy and competency program | Training programs to ensure personnel develop AI literacy, risk awareness, and operational competency | Eisenberg et al (2025) |
| Supporting external safety research | Actions by which firms support external safety-relevant researchers | Future of Life Institute (2024) |
| Testimonies to policymakers | Direct communication with policymakers about potential catastrophic risks from advanced AI | Future of Life Institute (2024) |
| Avoiding hype | AGI labs should avoid releasing models in ways likely to create hype (e.g., by overstating results) | Schuett et al (2023) |
| Company risk culture | Top management expressing values and providing guidance to employees on safety | Campos et al (2025) |
| Apply lifecycle risk controls | Implement risk-reduction controls throughout AI lifecycle (auditing, red-teaming, staged release) | Barrett et al (2024) |
| Trustworthy AI practices | Integrate characteristics of trustworthy AI into organizational policies and procedures | Barrett et al (2024) |
| Safety-first culture | Policies and practices to foster critical thinking and safety-first mindset to minimize negative impacts | Barrett et al (2024) |
| Interdisciplinary collaboration | Prioritize interdisciplinary AI actors with demographic diversity and broad domain expertise | Barrett et al (2024) |
| Sustain deployed system value | Mechanisms to sustain the value of deployed AI systems | Barrett et al (2024) |
| External input into tool development | Work with external actors to ensure safety tools are usable and suit their needs | UK DSIT (2023) |
| Cost-inducing training for malicious use | Design AI models to make post-training modifications too costly for malicious uses | Gipiškis et al (2024) |